\documentclass[%}
 reprint,
 longbibliography,
 amsmath,amssymb,
 aps,
]{revtex4-2}

\usepackage[pdftex,colorlinks=true,linkcolor=blue,bookmarks=false,citecolor=blue,urlcolor=blue,hypertexnames=false]{hyperref}
\usepackage{graphicx}% Include figure files
\usepackage{dcolumn}% Align table columns on decimal point
\usepackage{bm}% bold math
\usepackage{siunitx} % for nice unit handling
\usepackage{float}
\usepackage{xcolor}
\usepackage{hyperref}
\usepackage{cancel}
\usepackage{comment}

\newcommand{\ybk}[1]{\textcolor{black}{#1}}

\begin{document}

\title{Tunably-polarized driving light controls the phase diagram \\ of 1D quasicrystals and 2D quantum Hall matter}

\author{Yifei Bai}
\author{David M. Weld}
\email{weld@ucsb.edu}
 \affiliation{Department of Physics, University of California, Santa Barbara, California 93106, USA}
 
\begin{abstract}
The well-known mapping between 1D quasiperiodic systems and 2D integer quantum Hall matter can also be applied in the presence of driving.  Here we explore the effect of time-varying electric fields on the transport properties and phase diagram of Harper-Hofstadter materials. 
We consider light of arbitrary polarization illuminating a 2D electron gas at high magnetic field; this system maps to a 1D quasicrystal subjected to simultaneous phasonic and dipolar driving.
We show that this generalized driving generates a tessellated phase diagram featuring a nested duality-protected pattern of metal-insulator transitions. Circularly or elliptically polarized light can create an extended critical phase, opening up a new route to achieving wavefunction multifractality without fine-tuning to a critical point. We describe in detail a path to experimental realization of these phenomena using lattice-trapped ultracold atoms.

\end{abstract}

\maketitle

Quasiperiodic systems host a rich array of physical phenomena ranging from localization \cite{aubry1980analyticity, jitomirskaya_metal-insulator_1999_AMEq, Roati2008_nature, Lahini_AAexp_photonics_PhysRevLett.103.013901} to fractal spectra \cite{Hofstadter_OG_PhysRevB.14.2239, Jagannathan_Fib_RevModPhys.93.045001, jitomirskaya_metal-insulator_1999_AMEq} to criticality \cite{Evers_AL_RevModPhys.80.1355, Han_cri_bicrit_modifiedAAH_PhysRevB.50.11365, Desideri_CriModeExp_PhysRevLett.63.390, Zhou_AMEInteractionPhysRevLett.131.176401, Lin_generalToCP_PhysRevB.108.174206} to non-trivial topology \cite{Kraus_QP_pumpingExp_PhysRevLett.109.106402, Kraus_2D4DQH_PhysRevLett.111.226401, kraus_quasiperiodicity_2016, Prodan_virtual_QC_top_PhysRevB.91.245104, zilberberg_topology_2021}. Still richer possibilities arise in the presence of external driving, including a competition between thermalization and many-body localization~\cite{Bordia2017_FloquetMBL_Exp, KAAH_num_PhysRevB.106.054312} and potentially extended critical phases featuring multifractal wavefunctions~\cite{borgonovi_spectral_1995, ketzmerick1999efficient, prosen2001dimer, KAAH_num_PhysRevB.106.054312, shimasaki2022anomalous,Liu_modifiedAAH_PhysRevB.91.014108, Liu_LSP_10.21468/SciPostPhys.12.1.027, Roy_MFwithoutFineTune_10.21468/SciPostPhys.4.5.025, Wang_Raman-Critical_PhysRevLett.125.073204,Gonifmmode_CP_theory_PhysRevLett.131.186303, Gonifmmode_CP_theory_PhysRevLett.131.186303, Lin_generalToCP_PhysRevB.108.174206, XIAO_Critical_MomentumLattice_20212175, Li_modifiedAAH_exp_2023,Wang_MBC_PhysRevLett.126.080602}. 
\ybk{Identifying a simple experimental control knob that tunes and connects these dynamical phenomena is the open challenge we address in this theoretical manuscript.}

All these phenomena arise from the long-range spatial correlation of quasiperiodicity, which itself is the result of projecting a higher-dimensional periodic structure into a lower-dimensional physical space ~\cite{harper_single_1955, Kraus_QP_pumpingExp_PhysRevLett.109.106402, Kraus_2D4DQH_PhysRevLett.111.226401, kraus_quasiperiodicity_2016, Prodan_virtual_QC_top_PhysRevB.91.245104, zilberberg_topology_2021}. The connection to the higher-dimensional space also gives rise to a degree of freedom called a phason mode. A recent experiment~\cite{shimasaki2023reversible} illustrated a connection between two apparently unrelated localization phenomena --- dynamical and Aubry-Andr\'e localization --- by mapping a rapidly oscillating phasonic modulation to linearly polarized monochromatic light in the higher-dimensional space. This naturally raises the question of what the effects are of \emph{arbitrarily} tuning the polarization of irradiation in the higher-dimensional space.

In this work, we theoretically demonstrate that tunably polarized driving can be used to control localization, generate extended critical phases, and induce Floquet topological insulators by tuning the system along the axes of an intricately tesselated phase diagram in the space of drive polarizations. Specifically, we consider simultaneous dipolar \cite{lignier_dynamical_2007_DLExp} and phasonic  \cite{Rajagopal_phasonic_PhysRevLett.123.223201, shimasaki2023reversible} modulation in the 1D physical space. The phase difference between these two modulations maps to the polarization of applied light in the higher-dimensional space, enabling quantum simulation in a one-dimensional system of the effects of arbitrarily-polarized optical driving of a 2D electron gas at high magnetic field. We report the following main results. First, the combination of phasonic and dipolar modulation allows coherent control of both the tunneling \cite{lignier_dynamical_2007_DLExp} and quasi-disorder strengths \cite{shimasaki2023reversible}, resulting in a tessellated localization phase diagram with interlaced localized and delocalized phases. We note that this control toolset allows all elements of the non-interacting Hamiltonian to be flipped in sign, effectively reversing the direction of the flow of time. We demonstrate that circularly polarized illumination naturally and without fine-tuning generates an extended exotic phase hosting multifractal wavefunctions  \cite{Roy_MFwithoutFineTune_10.21468/SciPostPhys.4.5.025}. \ybk{Distinct from previous work on drive-induced extended criticality~\cite{shimasaki2022anomalous, Gonçalves_Ribeiro_Khaymovich_2023, Ray_Ghosh_Sinha_2018}, this result does not rely on mixing localized and ballistic modes, but is a direct consequence of the drive polarization.} 
Finally, we propose a straightforward experimental realization of all these phenomena using ultracold atoms in a driven bichromatic optical lattice. 

The paper is organized as follows. In Section \ref{sec:bg} we review the projection of quasiperiodic systems from higher-dimensional space. Readers familiar with the mapping can skip to Section \ref{sec:DHHtoDDAA} where we extend this mapping to include spatially homogeneous irradiation in the higher-dimensional space and comment on the effect of its polarization in the 1D space. We discuss the drive-induced tessellated localization phase diagram in Section \ref{sec:HFPD}, the effective time-reversal protocol in Section \ref{sec:time-reversal}, \ybk{and} Floquet-engineering of critical phases in Section \ref{sec:criticalPhase}.
A possible experimental realization with driven ultracold atoms is discussed in Section \ref{sec:exp}. We conclude by noting some intriguing future possible directions in Section \ref{sec:conclusion}.

%\onecolumngrid
\begin{figure*}[t]
    \centering
    \includegraphics[width = 0.95\textwidth]{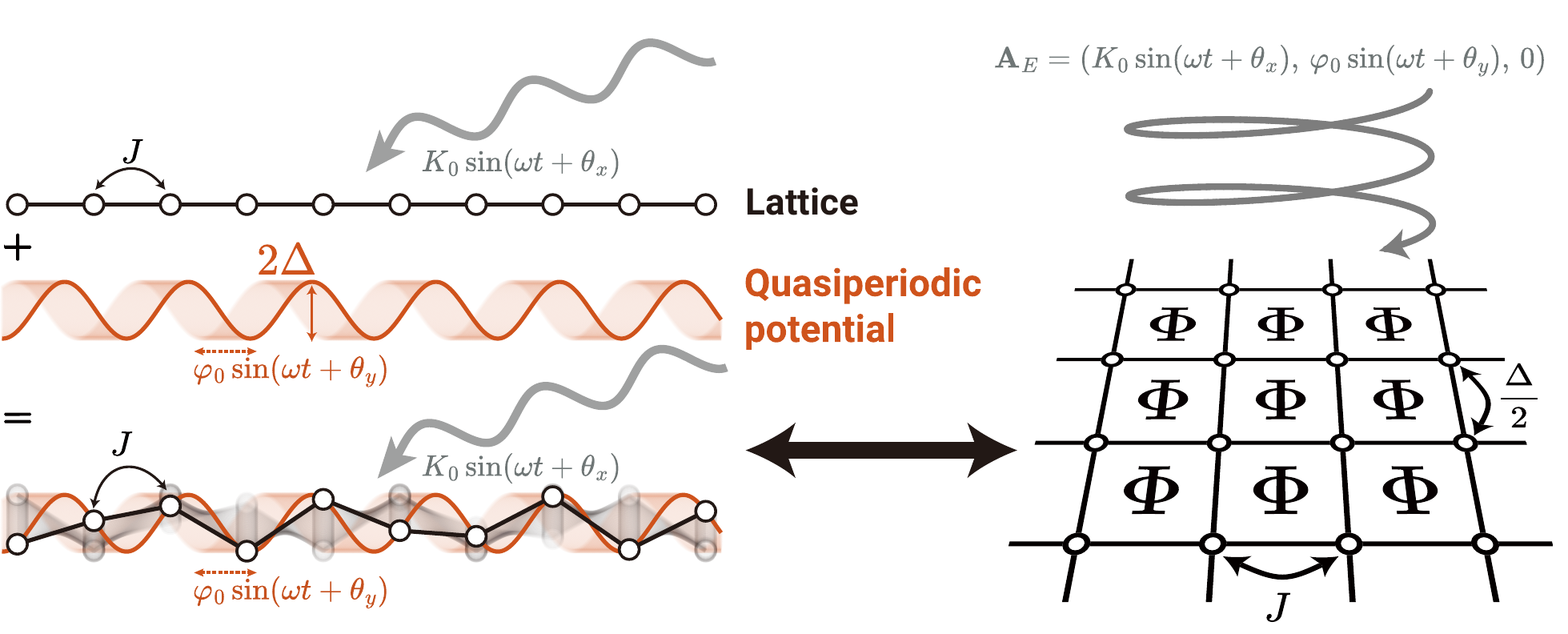}
    \caption{Diagram of the relation between the 1D driven quasicrystal  %DDAAH model (\ref{DAAH2}) 
    (left) and the 2D driven Harper-Hofstadter model (right). \textbf{(a)} The 1D quasicrystal is formed by a tight-binding lattice with nearest-neighbor (NN) hopping strength $J$ and a modulated on-site sinusoidal potential with strength $\Delta$ (orange) and a spatial period incommensurate to that of the lattice. The system is irradiated by the dipolar vector potential $K_0\sin(\omega t + \theta_x)$ (grey) and subject to phasonic modulation (shaded orange) with $\varphi_0\sin(\omega t + \theta_y)$. \textbf{(b)} The driven Harper-Hofstadter model is a 2D tight-binding square lattice with a magnetic flux per unit cell $\Phi := 2\pi\beta$, irradiated by light with the vector potential $\mathbf{A}_E$. $J$ is the NN tunneling strength along the horizontal direction and $\Delta/2$ becomes the tunneling strength along the vertical virtual direction. The dipolar and phasonic modulations in 1D map to the  components of the illuminating field $\mathbf{A}_E$ along the horizontal and vertical dimensions in 2D, respectively. }
    \label{fig:schematic}    
\end{figure*}
%\twocolumngrid

\section{Quasiperiodicity from higher dimensions} \label{sec:bg}
% introduce QC and connection to higher dimensional picture
Consider the paradigmatic Aubry-Andr\'e-Harper (AAH) model \cite{aubry1980analyticity}
\begin{align*}
    \hat H_{\mathrm{AAH}} (\kappa) = \sum_j -J\hat c_{j+1}^\dagger \hat c_j + \mathrm{h.c.} + \Delta  \cos(2\pi\beta j + \kappa) \hat n_j,
\end{align*}
where $J$ is the tunneling strength, $\Delta$ is the strength of the quasiperiodic potential, $\hat c^{(\dagger)}_j$ is the spinless annihilation (creation) operator at the site $j$, $\hat n_j := \hat c_j^\dagger \hat c_j$, and $\beta$ is the incommensurate ratio that defines the quasiperiodicity, \textit{i.e.} the ratio of the spatial period of the underlying lattice to that of the quasiperiodic modulation.

The AAH model supports a localization phase transition at $|\Delta/J| = 2$ when $\beta$ is irrational~\cite{aubry1980analyticity, Roati2008_nature, Lahini_AAexp_photonics_PhysRevLett.103.013901}. When $\Delta > 2J$, all eigenstates are localized and the system is in the localized phase; when $\Delta < 2J$, all eigenstates are extended and the system is in the delocalized phase. At the critical point, all eigenstates are multifractal with non-trivial scaling properties.

The parameter $\kappa$ describes a phase difference between the potential term and the underlying lattice. It does not affect the localization transition \cite{jitomirskaya_metal-insulator_1999_AMEq}, but encodes important information. When $\beta$ is rational, a change in $\kappa$ costs energy and the bulk spectrum of the system depends on $\kappa$. In contrast, when the system is aperiodic, the bulk spectrum is independent of $\kappa$, because there is a dense set of shifts in $[0, 2\pi)$ that leaves the bulk spectrum unchanged due to the irrationality of $\beta$ \cite{Kraus_QP_pumpingExp_PhysRevLett.109.106402, aubry1980analyticity, Aubry_1980, Aubry_1981}. The parameter $\kappa$ thus describes a zero-frequency shift for aperiodic systems, and for this reason it is called the phasonic degree of freedom \cite{Aubry_1980}. 

Interestingly, the AAH model can also be obtained as one of the Fourier components of the 2D Harper-Hofstadter (HH) model \cite{harper_single_1955, Kraus_QP_pumpingExp_PhysRevLett.109.106402}, and from this point of view $\kappa$ represents the quasimomentum along the extra dimension. To see this, we first extend the definition of the 1D operators into this dimension via the two-index operators $\hat c_{j, \kappa}^{(\dagger)}$ with anticommutation relation $\{\hat c_{j, \kappa}, \hat c_{j', \kappa'}^{\dagger}\} = \delta_{j,j'} \delta_{\kappa, \kappa'}$. We further define the following 2D annihilation (creation) operators $\hat d_{j, l}^{(\dagger)}$ by Fourier transform $\hat c_{j, \kappa}^{(\dagger)} = \sum_l e^{-i \kappa l}\hat d_{j, l}^{(\dagger)}$. Assuming a periodic boundary condition or infinite extent along the extra dimension, we obtain the 2D Harper-Hofstadter model
\begin{align*}
    \hat H_{\mathrm{HH}} &= \int^{2\pi}_0 \frac{d\kappa}{2\pi} \hat H_{\mathrm{AAH}} (\kappa) \\
    &= \sum_{j, l} -J \hat d_{j+1, l}^\dagger \hat d_{j,l} + \frac{\Delta}{2} e^{-i 2\pi\beta j} \hat d_{j, l+1}^\dagger \hat d_{j,l} + \mathrm{h.c.}. 
\end{align*}
The incommensurate ratio $\beta$ then represents the ratio of the magnetic flux over the flux quantum in the 2D picture, and the quasiperiodic strength $\Delta$ becomes the tunneling along the extra dimension. Historically, this mapping was first recognized by Harper \cite{harper_single_1955} and later explored in the context of the famous Hofstadter butterfly energy spectrum \cite{Hofstadter_OG_PhysRevB.14.2239}. 

\begin{figure*}[htb!]
    \centering
    \includegraphics[width = \linewidth]{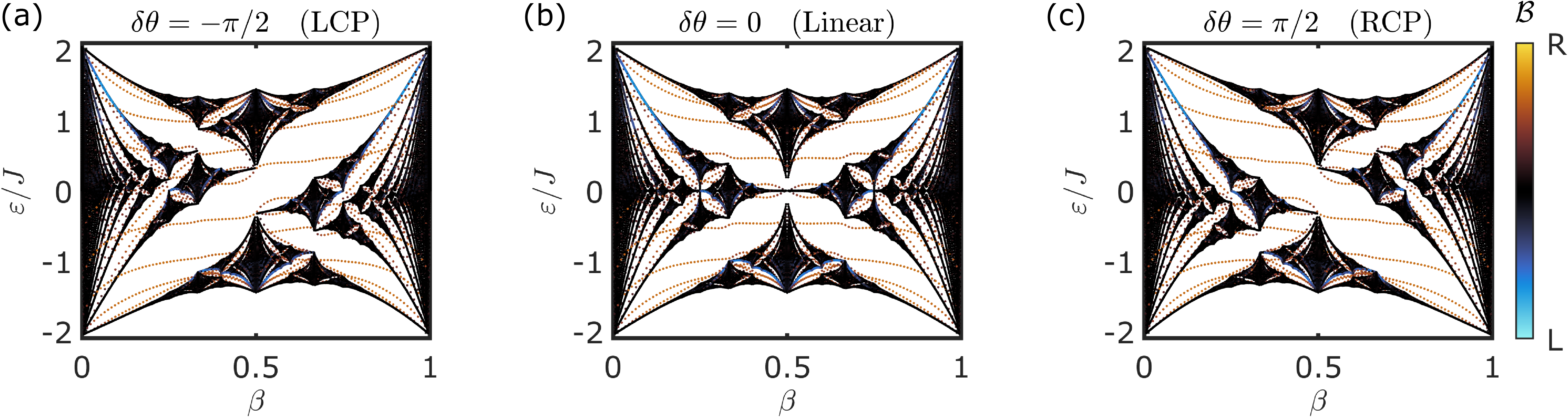}
    \caption{Quasienergy spectra of the DDAAH model for different drive polarizations: \textbf{(a)} left-handed circular polarization (LCP); \textbf{(b)}: linear polarization; and \textbf{(c)} right-handed circular polarization (RCP). The color represents the edge-locality marker $\mathcal{B}$ (see Eq.(\ref{eq:edgeLocMarker})) that identifies the edge states in the spectra localized at the right (R) or left (L) end. Here $\Delta/J = 2$, $K_0 = \varphi_0 = 1.5$, $\hbar \omega = 20J$, $\kappa = 0$ and $\theta_x = \pi/2$ with 100 lattice sites and 5 Floquet-Brillouin zones.}
    \label{fig:butterfly}    
\end{figure*}
\subsection{From irradiated Harper-Hofstadter Lattice to Doubly Driven AAH Quasicrystal} \label{sec:DHHtoDDAA}
We now consider the 2D HH model driven by homogeneous and monochromatic light irradiation incident from a direction perpendicular to the plane \cite{iomin_driven_1998, iomin_models_1999_DrivenAA, iomin_model_2000}. We choose a gauge where the scalar potential is zero, and the dimensionless vector potential is
\begin{align*}
    &\mathbf{A} = \mathbf{A}_B + \mathbf{A}_E, \quad \mathbf{A}_B = (0,2\pi\beta j, 0), \\
    & \mathbf{A}_E = \left(K_0 \sin (\omega t + \theta_x), \, \varphi_0 \sin (\omega t + \theta_y), \, 0 \right), 
\end{align*}
where we have explicitly separated the electric ($\mathbf{A}_E$) and magnetic ($\mathbf{A}_B$) contributions. $K_0$ and $\varphi_0$ are dimensionless field amplitudes along the $j$ and $l$ axis, respectively. The polarization of the irradiation is controlled by the phase difference $\delta \theta := \theta_y - \theta_x$ as follows: \begin{enumerate}
    \item When $\delta \theta = 0$ or $\pi$, the irradiation is linearly polarized. 
    \item When $\delta \theta = \pm \pi/2$, the irradiation is elliptically polarized. The semi-major and -minor axes of the polarization ellipse are parallel to the lattice axes. Additionally, the irradiation becomes circularly polarized when $K_0 = \varphi_0$. 
\end{enumerate}
For the irradiated system, we employ the Peierls substitution and the 2D driven Harper-Hofstadter model becomes \cite{zhao_floquet_2022}
% By virtue of Peierls substitution, this 2D driven HH model becomes \cite{zhao_floquet_2022}
\begin{align*}
    \hat H_\mathrm{DHH} = \sum_{jl} &-Je^{-iK_0 \sin (\omega t + \theta_x)} \hat d_{j+1, l}^\dagger \hat d_{j,l}  \\
    &+ \frac{\Delta}{2} e^{-i \left(2\pi \beta j+ \varphi_0 \sin (\omega t + \theta_y)\right)} \hat d_{j, l+1}^\dagger \hat d_{j,l} + \mathrm{h.c.}.
\end{align*}
The projection from the virtual 2D space into the physical 1D space can then be applied because the incident light is spatially homogeneous. This leads to the doubly-driven AAH (DDAAH) model
\begin{align}
    \hat H (t) = \sum_j &-J e^{-i K_0 \sin(\omega t + \theta_x)} \hat c_{j+1}^\dagger \hat c_j + \mathrm{h.c.} \nonumber \\
    & + \Delta \cos(2\pi\beta j + \kappa +  \varphi_0\sin(\omega t + \theta_y )) \hat n_j, \label{DAAH2}
\end{align}
where the $\kappa$ dependence in the 1D operators is dropped for convenience. A unitary transformation \cite{Eckardt_SFMI_PhysRevLett.95.260404} that takes the Hamiltonian (\ref{DAAH2}) into the length gauge helps elucidate its physical meaning:
\begin{align}
    \hat H (t) \rightarrow &\sum_j -J \hat c_{j+1}^\dagger \hat c_j + \mathrm{h.c.} + K\cos(\omega t+\theta_x ) j\hat n_j \nonumber \\
    & + \Delta \cos(2\pi\beta j + \kappa + \varphi_0\sin(\omega t + \theta_y ) ) \hat n_j , \label{DAAH1}
\end{align}
where $K := \hbar \omega K_0$ is the field amplitude in the 1D physical space. This Hamiltonian will be the central subject of this work. 

We see that dimensional reduction has mapped the field component parallel to the physical dimension (labeled by $j$) into a real oscillating field in 1D with amplitude $K$. We refer to this modulation as the dipolar modulation, since it has the form of an electric dipole force. The field along the extra dimension (labelled by $l$) is projected into a phasonic modulation \cite{shimasaki2023reversible} with amplitude $\varphi_0$. This correspondence of parameters between 1D and 2D physical models is schematically shown in Fig \ref{fig:schematic}. We can thus interpret a time-dependent phason as a virtual electric field along the extra dimension \cite{iomin_driven_1998, iomin_models_1999_DrivenAA, iomin_model_2000, kraus_quasiperiodicity_2016, Lohse2018_2D4DQH_Exp, shimasaki2023reversible}. This interpretation underpins, for example, topological pumping in the bichromatic lattice \cite{Kraus_QP_pumpingExp_PhysRevLett.109.106402, Lohse2018_2D4DQH_Exp}: adiabatic linear scanning of the phason maps to a static electric field (and thus charge-pumping) along the extra dimension in the higher-dimensional HH model. 
For our purposes, a crucial feature of this mapping is that the polarization of the radiation in the 2D space maps to the phase difference between the dipolar and phasonic modulations in 1D. In light of this, we will use the terminology of polarization to refer to these modulations for the remainder of the paper. 

As Fig.~\ref{fig:butterfly} demonstrates, the impact of different polarizations is evident in the quasienergy spectra as a function of $\beta$ (the Floquet-Hofstadter butterfly) \cite{zhao_floquet_2022, Kooi_genesis_PhysRevB.98.115124, Hatsugai_HH_NNN_spectrum_PhysRevB.42.8282}. We observe that linearly-polarized modulation preserves the structure of the Hofstadter butterfly (Fig. \ref{fig:butterfly}(b)), while circularly-polarized modulation asymmetrically distorts it (Fig. \ref{fig:butterfly}(a) and (c)). This is due to the different symmetry-breaking properties of these different polarizations of the incident light in the 2D space. Circularly polarized light breaks both time-reversal and sublattice symmetries in the higher-dimensional space; consequently, the corresponding symmetries in the Hofstadter butterfly spectrum (reflection symmetry about $\beta = 1/2$ and particle-hole symmetry about $E = 0$, respectively) are broken ~\cite{zhao_floquet_2022}. 
On the other hand, linearly polarized light does not break these symmetries \cite{dunlap_dynamic_1986_DL, zhao_floquet_2022}. Ref. \cite{zhao_floquet_2022} also noted other small differences of the Floquet-Hofstadter butterfly spectra compared to the time-averaged prediction. These differences, however, are not due to symmetry-breaking from the polarization, but deviations from the time-averaged description, which only strictly holds at $\omega \rightarrow \infty$.

Much of the physics of the 1D model (Eq.~(\ref{DAAH2})) can be understood in the 2D picture, even without any explicit calculation. First, time-periodic driving modifies the localization properties of the system. It is well-known that a lattice system under time-periodic driving exhibits dynamic localization \cite{dunlap_dynamic_1986_DL, lignier_dynamical_2007_DLExp}. This phenomenon can be intuitively understood as time-averaged Bloch oscillation in an AC electric field. Depending on the field strength projected to each dimension, the driving coherently re-scales the tunneling strength along that dimension, and can stop the tunneling of the particle along that dimension when the driving amplitude equals certain resonant values. This competition of tunneling suppression between different axes in the higher-dimensional space will impact the localization properties in the projected physical space, as the tunneling strength along the extra dimension becomes the strength of the quasiperiodic chemical potential variation through dimensional reduction. As we will see in Sections \ref{sec:HFPD} and \ref{sec:time-reversal}, this mechanism leads to a rich tessellated localization phase diagram in the space of drive polarizations, and allows for coherent and time-reversible control of localization dynamics. 

\ybk{Moreover,} driving with circularly-polarized light will in general lead to \ybk{effective} next-nearest-neighbor (NNN) hopping in the higher-dimensional space through virtual photon absorption and emission processes \cite{Kitagawa_PhysRevB.84.235108}. When projected to the 1D physical space, the NNN hopping can lead to mobility edges and critical states \cite{Han_cri_bicrit_modifiedAAH_PhysRevB.50.11365, kraus_quasiperiodicity_2016}. We show in Section \ref{sec:criticalPhase} that the NNN hopping in the 2D higher-dimensional space allows Floquet engineering of an extended critical phase hosting multifractal wave functions without fine-tuning. \ybk{This polarization-induced effective NNN hopping also modifies the topology of the system due to broken time-reversal symmetry. }
Given that our primary focus here is on localization phenomena in the driven quasiperiodic model, we will provide just a brief discussion of topology in Appendix \ref{sec:piHHtoSSH}, where we connect the driven Harper-Hofstadter model to the generalized Rice-Mele and Su–Schrieffer–Heeger models. 

\subsection{High-frequency expansions}
In this study, we will focus on the high-frequency regime $(\hbar \omega \gg J, \Delta)$ such that there is no overlap between Floquet bands~\cite{Bukov_universal_high_freq, Eckardt_2015_HFE}. We leave exploration of the additional richness of the low-frequency \cite{Rodriguez-Vega_2018} and/or resonant regimes \cite{Goldman_resonantFreq_PhysRevA.91.033632} for future work. We compute the Fourier expansion of $\hat H (t) = \sum_{m=-\infty}^\infty \hat H_m e^{im\omega t}$ and derive the effective Hamiltonian using the high-frequency expansion ~\cite{Bukov_universal_high_freq, Goldman_PhysRevX.4.031027, Eckardt_2015_HFE} up to the third order
\begin{equation}
    \hat H_\mathrm{eff} =\sum_{n=0}^{\infty} \hat H^{(n)} =\hat H^{(0)} + \hat H^{(1)} + \hat H^{(2)} + \mathcal{O}\left((\hbar\omega)^{-3} \right) \label{eq:HFE}
\end{equation}
The $(n+1)^\mathrm{th}$ order term $\hat H^{(n)}$ in the expansion will scale as $1/\omega^n$, so in the high-frequency regime terms with $n \ge 2$ will be small perturbations compared to the $n = 0,1$ terms. Nonetheless, the contributions of these higher-order terms can be important and can be clearly observed in some regions of the phase diagram, as we will see shortly. 

Terms in the high-frequency expansion can be computed using the Fourier coefficients $\hat H_m$ of the time-periodic Hamiltonian. For the first two terms \cite{Bukov_universal_high_freq, Goldman_PhysRevX.4.031027, Eckardt_2015_HFE}, 
\[
\hat H^{(0)} = \frac{1}{T}\int^{T/2}_{-T/2}\mathrm{d}t \, \hat H(t), \quad \hat H^{(1)} = \sum_{m=1}^\infty \frac{\left[ \hat H_m ,\hat H_{-m}\right]}{m\hbar \omega}.
\]
The first term $\hat H^{(0)}$ is just the time-averaged Hamiltonian over a period, and $\hat H^{(1)}$ is a sum over virtual $m-$photon absorption and emission processes \cite{Kitagawa_PhysRevB.84.235108}. In practice, the series is summed up to some finite Fourier index $m_\mathrm{cutoff}$. In particular, for linearly polarized modulations, $\hat H_m = \hat H_{-m}$ and so $\hat H^{(1)}$ vanishes.

\begin{figure*}[htb!]
    \centering
    \includegraphics[width = \linewidth]{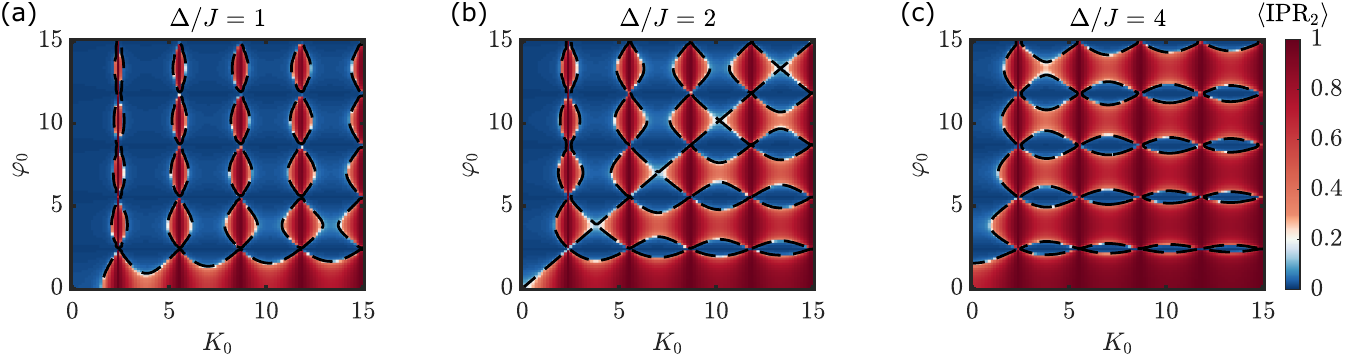}
    \caption{The tessellated phase diagrams of $\hat H_\mathrm{eff} \approx \hat H^{(0)}$ at \textbf{(a)} $\Delta/J = 1$ (delocalized phase without driving), \textbf{(b)} $\Delta/J = 2$ (critical point when no driving), and \textbf{(c)} $\Delta/J = 4$ (localized phase when no driving), computed by diagonalization of the Floquet block Hamiltonian over 5 Floquet-Brillouin zones with $\theta_x = \pi/2$ and $\theta_y = 0$ and averaged over 10 different values of $\kappa$. The dashed curves are phase boundaries predicted by Eq. \ref{eq:AA_HF}. The system size is $L = 100$ sites and $\hbar \omega = 100J$ with open boundary conditions. }
    \label{fig:cascadeLocTransition}    
\end{figure*}

\section{Floquet-Engineering of Localization Transition and Critical Phases}
We are now in a position to investigate the localization phase diagram of the DDAAH Hamiltonian (\ref{DAAH2}). We will assume $\beta = (\sqrt{5}-1)/2$ for the numerical studies in this section, although the results are applicable for any irrational $\beta$. 

\begin{figure}[h!]
    \centering
    \includegraphics[width = \linewidth]{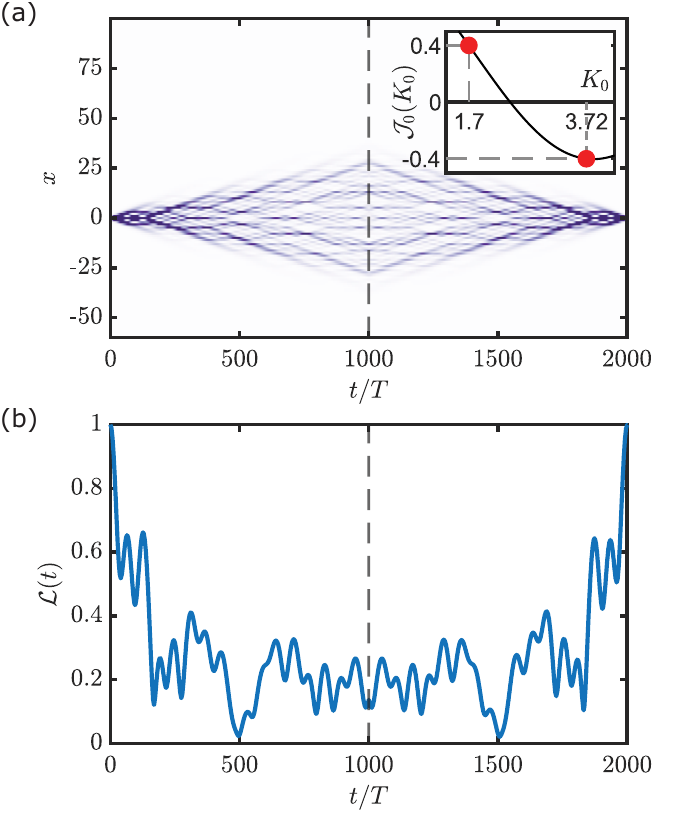}
    \caption{Controlling effective time-reversal in the DDAAH model. The system is driven for 1000 cycles for forward propagation at $K_0^\mathrm{for} \approx 1.6965$ ($\mathcal{J}_0(K_0^\mathrm{for}) = 0.4$) and another 1000 cycles for backward propagation at $K_0^\mathrm{back} \approx 3.7152$ ($\mathcal{J}_0(K_0^\mathrm{back}) = -0.4$), shown as the red dots in the inset of \textbf{(a)}. The time-reversal dynamics are observed in both \textbf{(a)} the time-evolution of the density profile and \textbf{(b)} the return probability $\mathcal{L}(t)$. The initial state is a single-site excitation $\psi(t) = \delta_{j,0}$, with parameters $\Delta/J = 1$, $\theta_x = \theta_y = 0$ and $\hbar \omega = 100$. The Hamiltonian (Eq.\ref{DAAH1}) is numerically integrated with the second order split-operator method with a time step of $dt \approx 1.26\times 10^{-5} \hbar/J$.} 
    \label{fig:time-reversal}    
\end{figure}

\subsection{Tessellated phase diagram} \label{sec:HFPD}
When the driving frequency is large enough compared to all other energy scales of the system, only the time-averaged term $\hat H^{(0)}$ remains in the effective Hamiltonian: 
\begin{align*}
    \hat H_\mathrm{eff} &\approx \hat H^{(0)} \\
    = &\sum_j -J_\mathrm{eff} \left(\hat c_{j+1}^\dagger \hat c_j + \mathrm{h.c.}\right) + \Delta_\mathrm{eff}\cos \left(2\pi\beta j + \kappa \right)\hat n_j,
\end{align*}
where $J_\mathrm{eff} := J \mathcal{J}_0 (K_0)$ and $\Delta_\mathrm{eff} := \Delta \mathcal{J}_0 (\varphi_0)$. $\hat H_\mathrm{eff}$ has the form of the static AAH Hamiltonian, but with both the effective tunneling strength and the effective quasiperiodic potential strength re-scaled by a zeroth order Bessel function. The re-scaling of the tunneling strength leads to dynamic localization~\cite{dunlap_dynamic_1986_DL, lignier_dynamical_2007_DLExp} when $K_0$ is tuned to a zero of $\mathcal{J}_0 (K_0)$. The re-scaling of the quasiperiodic potential strength due to the phasonic modulation leads to destruction of localization when $\varphi_0$ is tuned to a zero of $\mathcal{J}_0$, a phenomenon which was only recently experimentally explored ~\cite{shimasaki2023reversible}. The polarization plays no role in this time-averaged case.

Despite their apparent differences, the underlying mathematical origins of these two phenomena (dynamic localization due to dipolar driving and dynamic delocalization due to phasonic driving) are closely related. We first discuss them in the 1D physical space. Dynamic localization occurs when the dispersion time-averages to zero for every quasimomentum \cite{Dunlap_Kenkre_1988_DLmomentumSpace}, so that the propagation in position space is bounded. Specifically, within one driving period, the dynamical phases of each quasimomentum basis wind forward for the first half of the cycle and backward for the remainder of the cycle, such that the time-averaged dispersion is zero for every quasimomentum. 

For phasonic destruction of localization, an identical picture arises but in the position basis. During the first half of the phasonic cycle, the phase accumulates differently at each lattice site due to the quasiperiodic potential. This phase winds back for the remainder of the modulation. The destruction of localization occurs when the phase winding at each lattice site due to the quasiperiodic potential time-averages to zero. When this cancellation happens, the wavepacket propagates as if there were no on-site modulation of the potential at all~\cite{shimasaki2023reversible}. 

The connection between dynamic localization and dynamic delocalization arises naturally in the 2D picture as well, where the phasonic modulation corresponds to irradiation polarized along the extra dimension. When $\varphi_0$ is tuned to a zero of $\mathcal{J}_0(\varphi_0)$, dynamic \emph{localization} occurs along the extra dimension; consequently, the 2D square lattice breaks into a set of disconnected 1D chains, each of which cannot support localization. Thus dynamic localization in the extra dimension leads to delocalization in the transverse dimension. 

In the presence of both types of modulation, there is a competition between dynamic localization  from dipolar modulation and dynamic delocalization from phasonic modulation. In the 2D picture this can be seen as an anisotropic effect of dynamic localization when the driving field is not aligned to the lattice axes~\cite{dunlap_dynamic_1986_DL}. This competition leads to an intricate duality-protected pattern of localization quantum phase transitions coherently controlled by the phasonic and dipolar driving amplitudes. The Aubry-Andr\'e duality argument predicts that these transitions occur at
\begin{equation}
    \bigg\vert\frac{\Delta_\mathrm{eff}}{J_\mathrm{eff}} \bigg\vert = \bigg \vert \frac{\Delta}{J}\frac{\mathcal{J}_0 (\varphi_0)}{\mathcal{J}_0 (K_0)} \bigg \vert = 2. \label{eq:AA_HF}    
\end{equation}

To probe the phase diagram, we numerically diagonalize the Hamiltonian (\ref{DAAH2}) to obtain the Floquet eigenstates (see Appendix \ref{apd:FlqAnalysis} for more details). Our diagnostic of localization is the inverse participation ratio of the Floquet eigenstates in the position basis,  defined as
\begin{equation}
    \mathrm{IPR}_2^{(i)} = \frac{\sum_{j=1}^L \vert \psi_j^{(i)} \vert^4}{ \left(\sum_{j=1}^L \vert \psi_j ^{(i)} \vert^2 \right)^2}
\end{equation}
where $L$ is the system size and $\psi_j^{(i)}$ is the $i^\mathrm{th}$ Floquet eigenstate in the position basis. A (non)vanishing $\mathrm{IPR}_2^{(i)}$ indicates a spatially extended (localized) state. We further define $\langle \mathrm{IPR}_2\rangle$, obtained by averaging the $\mathrm{IPR}_2^{(i)}$ over all Floquet eigenstates in the central Floquet-Brillouin zone which is defined by $-\hbar\omega/2 \leq \varepsilon \leq \hbar \omega/2$. The results are shown in Fig.~\ref{fig:cascadeLocTransition}. The phase boundaries predicted by the analytical high-frequency approximation (Eq. \ref{eq:AA_HF}) are shown as dashed curves and demonstrate excellent agreement with numerical results. The phase diagram has a tessellated self-dual structure with interlacing areas of localized and delocalized phases due to the competition between $\Delta_\mathrm{eff}$ and $J_\mathrm{eff}$. We emphasize that this rich localization phase diagram characterizes, and could be observed in, both a 1D quasicrystal subjected to combined dipolar and phasonic modulation, and a 2D electron gas at high magnetic field and strong irradiation \cite{Paul_directional_PhysRevLett.132.246402}.

\subsection{Effective time-reversal dynamics} \label{sec:time-reversal}
These results demonstrate that the localization quantum phase transition of the 1D DDAAH model can be coherently controlled by combined dipolar and phasonic modulation. In contrast to previous schemes where either the dipolar or the phasonic modulation is individually applied \cite{lignier_dynamical_2007_DLExp, shimasaki2023reversible}, the use of combined modulation allows for simultaneous control of the tunneling strength and quasiperiodic potential together, opening up new possibilities.  As an example of such a possibility, we describe a driving protocol that achieves effective time-reversal dynamics in the DDAAH model. 
\ybk{This is potentially of practical as well as intrinsic interest; the ability to effectively reverse time evolution is a key component of schemes for  experimentally probing fidelity decay or scrambling through Loschmidt echo or out-of-time-order correlators \cite{Yin_Loschmidt_loc_PhysRevA.97.033624}}. 

% As an example, we describe a driving protocol that achieves effective time-reversal dynamics in the DDAAH model. 
For simplicity, we consider both modulations to have the same amplitude ($\varphi_0 = K_0$). By simply choosing a sequence of two modulation amplitudes arranged around the zeros of the Bessel function $\mathcal{J}_0(K_0)$ such that $\mathcal{J}_0(K_0^\mathrm{for}) = -\mathcal{J}_0(K_0^\mathrm{back})$, the signs of all parameters in the effective Hamiltonian $\hat H^{(0)}$ are flipped, effectively reversing the direction of the flow of time. Results of numerical integration of the discrete time-dependent Schr\"odinger equation of the Hamiltonian (\ref{DAAH1}) are shown in figure \ref{fig:time-reversal}, for parameters and hold times well within the range of experimental feasibility. The time reversal is evident from both the wave function propagation and the return probability, defined as $\mathcal{L}(t) := \vert\langle \psi(t) \vert \psi(0)\rangle \vert^2$. 

\subsection{Floquet engineering of multifractality without fine-tuning} \label{sec:criticalPhase}

\begin{figure*}[htb!]
    \centering
    \includegraphics[width = \linewidth]{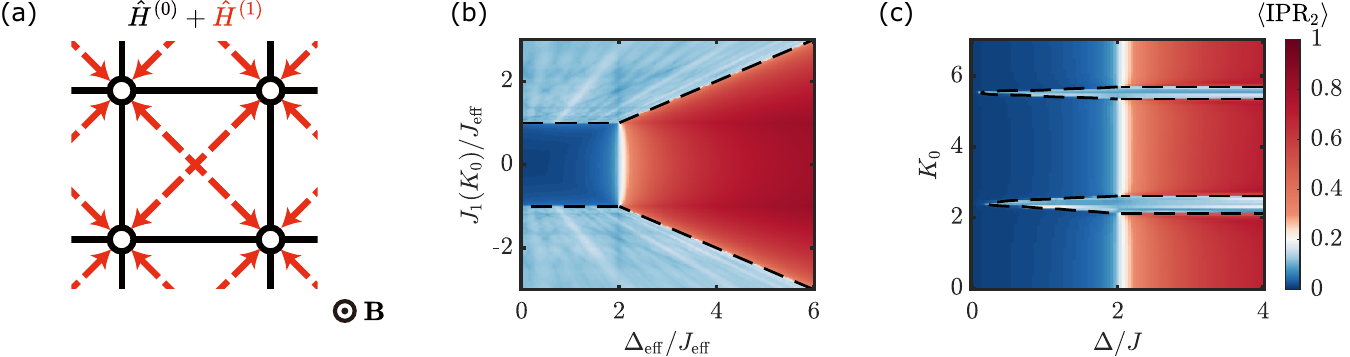}
    \caption{Floquet engineering a critical phase using circularly polarized modulation. \textbf{(a)} The effective Hamiltonian under circularly polarized light in the higher-dimensional space. The irradiation produces NNN hopping described by $\hat H^{(1)}$ (shown by the red arrows) on top of the underlying Harper-Hofstadter model. 
    \textbf{(b)} Phase diagram of the effective Hamiltonian $\hat H^{(0)} + \hat H^{(1)}$ showing $\langle\mathrm{IPR}_2 \rangle$ as a function of effective quasiperiodic potential strength $\Delta_\mathrm{eff}/J_\mathrm{eff}$ and off-diagonal quasiperiodic tunneling strength $J_1/J_\mathrm{eff}$. Three regions are observed: an extended phase (dark blue), a localized phase (red), and a critical phase (pale blue). \textbf{(c)} Phase diagram for the case of circularly polarized modulation, as a function of physical quasiperiodic potential strength $\Delta/J$ and dipolar modulation amplitude $K_0$, calculated using the Floquet block Hamiltonian $\mathcal{H}$ (\ref{eq:flq_blockH}). The critical phase emerges around the two zeros of $\mathcal{J}_0(K_0)$. Both \textbf{(b)} and \textbf{(c)} are averaged over 10 randomly selected values of $\kappa$. For \textbf{(c)}, 15 Floquet-Brillouin zones were included with $\delta \theta = -\pi/2$. The dashed curves are phase boundaries predicted by Eq. \ref{eq:CP_Condition}, which show excellent agreement with numerics. The system size is $L = 100$ sites with open boundary condition, $\theta_x = \pi/2$ and $\hbar \omega = 15J$.  }
    \label{fig:criticalPhase}    
\end{figure*}

In this section, we will show that the control capabilities afforded by tuning the drive polarization in the DDAAH model (\ref{DAAH1}) allow preparation of an extended critical phase of matter hosting multifractal wavefunctions without fine-tuning \cite{Han_cri_bicrit_modifiedAAH_PhysRevB.50.11365, Zhou_AMEInteractionPhysRevLett.131.176401, Lin_generalToCP_PhysRevB.108.174206}. The possibility of an extended critical phase is of interest from both theoretical and experimental perspectives; usually, multifractal wave functions only appear when a system is finely tuned to a phase boundary or mobility edge, which hinders experimental study. 

Previous work on irradiated graphene has shown that elliptically polarized irradiation introduces NNN hopping in the effective Hamiltonian through absorption and emission of virtual photons \cite{Kitagawa_PhysRevB.84.235108}. Moreover, when these NNN hoppings are present in the static 2D Harper-Hofstadter model with an incommensurate flux, they are known to introduce critical phases~\cite{Han_cri_bicrit_modifiedAAH_PhysRevB.50.11365, Avila2017_proof_CP}. As we will explicitly demonstrate below, it is therefore possible to engineer critical phases by applying elliptical irradiation to the Harper-Hofstadter Hamiltonian in the 2D space (see Fig. \ref{fig:criticalPhase}(a)). %The resulting projected 1D system can then host critical phases.

Motivated by this observation, we apply elliptical modulation to the 1D physical system. For simplicity, here we consider only the special case of circularly polarized modulation ($\delta \theta = \pm \pi/2$ and $K_0 = \varphi_0$), and defer the discussion of general polarization to Appendix \ref{apd:HFE_generic}. Then $\hat H^{(1)} \neq 0$ and the effective Hamiltonian is $\hat H_\mathrm{eff} \approx \hat H^{(0)}+\hat H^{(1)}$, where 
\[
\hat H^{(1)} = \sum_j J_1(K_0)\cos \left(2\pi\beta j + \pi\beta + \kappa \right)\hat c_{j+1}^\dagger \hat c_j+ \mathrm{h.c.}.
\]
$\hat H^{(1)}$ thus describes additional quasiperiodically modulated nearest-neighbor tunneling. The strength of this tunneling $J_1(K_0)$ is given by
\begin{equation}
    J_1(K_0) = \pm \frac{4J \Delta}{\hbar \omega}\sin(\pi\beta) \sum_{n=0}^\infty \frac{(-1)^{n+1}}{2n+1} \mathcal{J}_{2n+1}^2(K_0). \label{eq:j1}
\end{equation}
The sign of $J_1$ depends on whether the polarization is left-handed ($-$) or right-handed ($+$). 

The effective Hamiltonian $\hat H_\mathrm{eff}$ is an extended AAH model that includes a quasiperiodically modulated tunneling strength, which can be projected from the 2D static extended Harper-Hofstadter model with additional NNN hopping  \cite{Han_cri_bicrit_modifiedAAH_PhysRevB.50.11365, Chang_MFanalysis_modifiedAAH_PhysRevB.55.12971, drese_modified_AAH_1997, liu_localization_2015}. 
It features one of the best-known examples of a critical phase \cite{Han_cri_bicrit_modifiedAAH_PhysRevB.50.11365, Chang_MFanalysis_modifiedAAH_PhysRevB.55.12971, drese_modified_AAH_1997, liu_localization_2015}, whose existence in the thermodynamic limit has been proved in \cite{Avila2017_proof_CP}. % This is in contrast to the kicked AAH model \cite{borgonovi_spectral_1995, ketzmerick1999efficient, prosen2001dimer, KAAH_num_PhysRevB.106.054312, shimasaki2022anomalous}, for example, where the broad critical region is predicted to shrink as the system size increases \cite{ketzmerick1999efficient, prosen2001dimer}, albeit extremely slowly \cite{ketzmerick1999efficient, shimasaki2022anomalous}. 
The critical phase can be observed in the phase diagram of this extended AAH model shown in Fig. \ref{fig:criticalPhase}(b). When the extended AAH model is in the critical phase, wavefunctions of all eigenstates are multifractal \cite{Chang_MFanalysis_modifiedAAH_PhysRevB.55.12971, Avila2017_proof_CP} and a generalized AAH self-duality holds \cite{drese_modified_AAH_1997, Avila2017_proof_CP}. The boundary of the critical phase is given by
\begin{equation}
    |J_1(K_0)| > |J_\mathrm{eff}(K_0)| \text{ and } |J_1(K_0)| > \frac{1}{2}|\Delta_\mathrm{eff}(K_0)|. \label{eq:CP_Condition}
\end{equation} 
We thus expect the critical phase to emerge around the zeros of $\mathcal{J}_0 (K_0)$ where the ratio $|J_1(K_0) / J_\mathrm{eff}(K_0)| \rightarrow \infty$. Indeed, numerical diagonalization based on the Floquet block Hamiltonian (Eq. \ref{eq:flq_blockH}) shows two critical ``stripes" in the parameter space around the zeros of $\mathcal{J}_0 (K_0)$, displayed in Fig.\ref{fig:criticalPhase}(c). 

\begin{figure}[htb!]
    \centering
    \includegraphics[width = \linewidth]{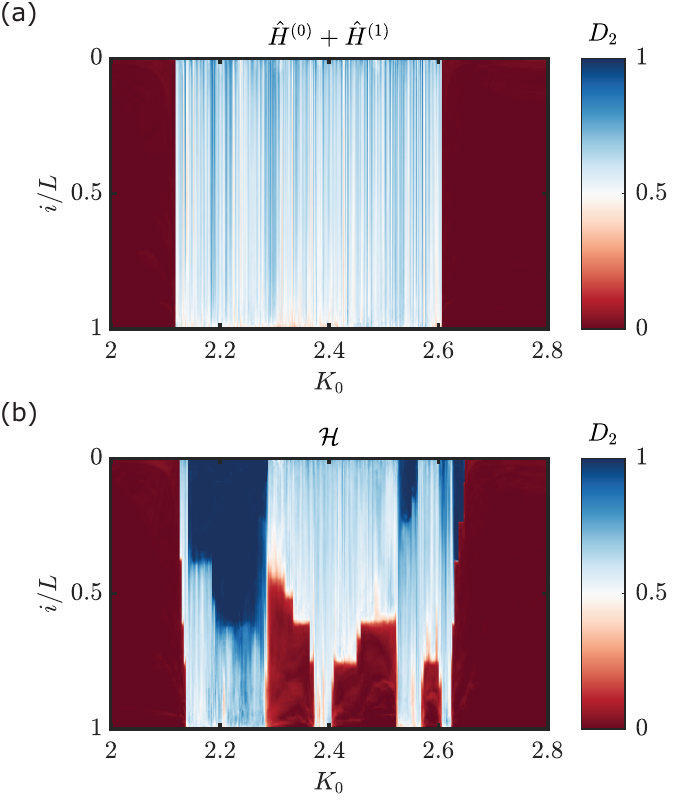}
    \caption{Eigenstate fractal dimensions $D_2$ for varying $K_0$ at a fixed quasiperiodic strength $\Delta/J = 3$. \textbf{(a)} Fractal dimension using the effective Hamiltonian $\hat H = \hat H^{(0)} + \hat H^{(1)}$ through one of the critical stripes. All eigenstates are critical inside the stripe. The system size is varied from $300$ to $6000$ sites. \textbf{(b)} Fractal dimensions using the Floquet block Hamiltonian with 5 Floquet-Brillouin zones. Some eigenstates are no longer multifractal. The system size is varied from $300$ to $3300$ sites due to increased computational cost. Colorbar shows $D_2$ at a driving frequency $\hbar \omega = 15J$ and $\kappa = 0$. Both calculations are done with open boundary conditions.} 
    \label{fig:MF_D2}    
\end{figure}

To demonstrate the multifractal nature of the Floquet eigenstates in the critical stripes, we calculate the fractal dimension $D_2$ of each Floquet eigenstate of the effective Hamiltonian up to $n=1$ ($\hat H_\mathrm{eff} \approx \hat H^{(0)}+\hat H^{(1)}$), defined as
\begin{equation*}
    \mathrm{IPR}_2^{(i)} \sim L^{-D_2},
\end{equation*}
where $L$ is the system size. For localized (extended) states, $D_2 = 0 \, (1)$, and for states with multifractal wave functions, $0 < D_2 < 1$. We focus on the first stripe of critical phase in Fig.\ref{fig:criticalPhase}(c) at $\Delta/J = 3$ and $2<K_0<2.8$. The result shown in Fig. \ref{fig:MF_D2}(a) demonstrates that for this effective Hamiltonian, all Floquet eigenstates in the critical stripe are indeed multifractal with non-trivial values of $D_2$.  % up to the $n=1$ term in the HFE. 

An important question in this context concerns the role of higher-order terms ($\hat H^{(n\geq 2)}$) in the high-frequency expansion. These terms, though small in the high-frequency regime, can potentially remove the multifractality of certain Floquet eigenstates in the spectrum, partly because these terms can break the delicate generalized self-duality of the extended AAH model \cite{drese_modified_AAH_1997, Avila2017_proof_CP}. To examine their impact, we numerically calculate the fractal dimension $D_2$ of the eigenstates of the Floquet block Hamiltonian instead of the effective Hamiltonian from the high-frequency expansion, thereby including every term in the high-frequency expansion in principle. The results, which appear in Fig. \ref{fig:MF_D2}(b), show that although certain Floquet eigenstates cease to be multifractal when the full Floquet Hamiltonian is considered, a finite fraction of Floquet eigenstates still exhibit multifractal characteristics with $D_2 \neq 0,1$. We further corroborate this result by numerical diagonalization of the Hamiltonian $\hat H^{(0)} + \hat H^{(1)} + \hat H^{(2)}$, which is discussed in the Appendix \ref{apd:D2Flq}. \ybk{We note, however, that because $\hat H^{(1)}$ is still the dominant contribution, quantities like the averaged IPR still serve as a meaningful metric despite the small mixing of the spectrum. }

We thus establish the existence of multifractality without fine-tuning in the DDAAH model when the combined modulations are tuned to correspond to circularly polarized light in the 2D space, and we further generalize this result to elliptical modulation in the Appendix \ref{apd:HFE_generic}. While previously proposed methods for generating critical phases depend upon mixing localized and extended eigenstates of the undriven system using low-frequency periodic driving \cite{Roy_MFwithoutFineTune_10.21468/SciPostPhys.4.5.025} or unbounded quasiperiodic potentials \cite{Liu_LSP_10.21468/SciPostPhys.12.1.027}, our model contains neither of those elements, and is inherently in the high-frequency regime, which should significantly alleviate the problem of interband heating in experimental platforms based on ultracold atoms \cite{shimasaki2022anomalous}. The effective Hamiltonian up to $\hat H^{(1)}$ corresponds to a model that had eluded direct experimental realization until very recently, with programmable superconducting qubits \cite{Li_modifiedAAH_exp_2023} and a constrained system size. We expect that the approach to achieving an extended fractal phase described in this section may significantly simplify experimental realization, as further discussed in section \ref{sec:exp}.

\section{Experimental realization} \label{sec:exp}
Finally, we discuss the experimental protocol needed to realize the DDAAH model and all phenomena discussed in the preceding section. Compared to some proposed and existing experimental realizations \cite{Wang_Raman-Critical_PhysRevLett.125.073204, Li_modifiedAAH_exp_2023, XIAO_Critical_MomentumLattice_20212175}, our model is relatively simple, only requiring already-demonstrated techniques in platforms such as photonic lattices \cite{Longhi_DLexp_PhysRevLett.96.243901, Lahini_AAexp_photonics_PhysRevLett.103.013901} and ultracold atoms \cite{lignier_dynamical_2007_DLExp, Roati2008_nature,shimasaki2023reversible}. We therefore expect the DDAAH Hamiltonian to be realizable in a wide range of experiments. 

We focus on the implementation of the DDAAH model using ultracold atoms in a shaken 1D bichromatic optical lattice \cite{Rajagopal_phasonic_PhysRevLett.123.223201, shimasaki2023reversible, Roati2008_nature}, which combines the AAH model and periodic modulation: 
\begin{align*}
    \hat H_\mathrm{BOL} (t) = \frac{\hat p^2}{2M} &+ V_P \cos \big ( 2k_P (\hat x - x_P(t)) \big) \\
    &+ V_S \cos \big(2k_S (\hat x - x_S(t) )+ \kappa \big).
\end{align*}
Here $M$ refers to the mass of the atoms, $k_{P,S}$ denote the wave vector of the primary and secondary lattices, respectively, and $\kappa$ accounts for any phase difference between the two lattices when there is no modulation. $V_{P,S}$ denote the lattice depths of the primary and secondary lattice, respectively. The Aubry-Andr\'e regime is achieved with $V_P \gg V_S$ and a deep primary lattice depth $V_P \gtrsim 10  E_\mathrm{r,P}$ \cite{Boers_Halthaus_PhysRevA.75.063404, Li_SDS_PhysRevB.96.085119} where $ E_\mathrm{r,P} = \hbar^2 k_P^2/(2M)$ is the recoil energy of the primary lattice, and the incommensurate ratio is the ratio of the lattice wave vectors, $\beta = k_S / k_P$. In this way, the weak secondary lattice will only shift the depth of each lattice site formed by the deep primary lattice and will not significantly alter the nearest-neighbor tunneling strength  \cite{Modugno_AA_parameter_2009}. $x_{P, S}(t)$ denote the position modulation (shaking) of the primary and secondary lattices, respectively. Although non-sinusoidal modulations are an interesting topic for future work, here we consider only sinusoidal modulation: 
\[
x_P (t) = \alpha_P\sin (\omega t + \theta_P), \quad x_S (t) = \alpha_{S}\sin (\omega t + \theta_S)
\]
where $\alpha_{P,S}$ are the shaking amplitudes of the primary and secondary lattices, respectively, and $\theta_{P,S}$ are the associated phases of each modulation. 

Because the deep primary lattice defines the lattice sites, we transform to the co-moving frame of the primary lattice \cite{Holthaus_floquet_lattice_2016}, which leads to
\begin{align*}
    \hat H_\mathrm{cm} (t) = &\frac{\hat p^2}{2M} + V_P \cos \big ( 2k_P \hat x \big) \\
    &+ V_S \cos \big(2k_S (\hat x - x_\mathrm{phason}(t) ) + \kappa \big) + \hat x F(t).
\end{align*}
The frame transformation gives rise to an inertial force $F(t) = M \ddot x_P (t)$, a dipolar modulation. The phasonic modulation $x_\mathrm{phason} (t) := x_S(t) - x_P (t) = \alpha_\mathrm{phason} \sin(\omega t + \theta_\mathrm{phason})$ is formed by the interference of $x_P (t)$ and $x_S (t)$. The amplitude and phase of the phasonic modulation are thus determined by 
\begin{align*}
    \alpha_\mathrm{phason} &= \sqrt{\alpha_P^2 + \alpha_S^2 - 2 \alpha_P \alpha_S \cos(\theta_S - \theta_P)}, \\
    \theta_\mathrm{phason} &= \operatorname{atan2}(u,v),
\end{align*}
where $u := {\alpha_S \sin \theta_S - \alpha_P \sin \theta_P}$, $v := \alpha_S \cos \theta_S - \alpha_P \cos \theta_P$ and $\operatorname{atan2}(u,v)$ is the 2-argument arctangent function. Note that the phasonic modulation amplitude $\alpha_\mathrm{phason}$ is also determined by the relative phase difference between position modulations of primary and secondary lattices. 

In the tight-binding regime achieved for large $V_P$, the Hamiltonian $\hat H_\mathrm{cm} (t)$ is equivalent to the DDAAH model (Eq.(\ref{DAAH2})) \cite{Modugno_AA_parameter_2009} with  dipolar and phasonic modulation amplitudes \cite{lignier_dynamical_2007_DLExp, shimasaki2023reversible}
\begin{equation}
     K_0 = \frac{\pi}{\hbar k_P} M\omega \alpha_P,\quad \varphi_0 = 2k_S \alpha_\mathrm{phason} \label{dipAndPhasonAmpValue}
\end{equation}
% K = \frac{\pi}{k_P} M\omega^2 \alpha_P $
and phases
\[
\theta_x = \theta_P + \frac{\pi}{2}, \quad \theta_y = \theta_\mathrm{phason}.
\]
The polarization is then controlled by the phase difference $\delta\theta = \theta_y - \theta_x$ between the dipolar and phasonic modulation. 

\begin{figure}[thb!]
    \centering
    \includegraphics[width = \linewidth]{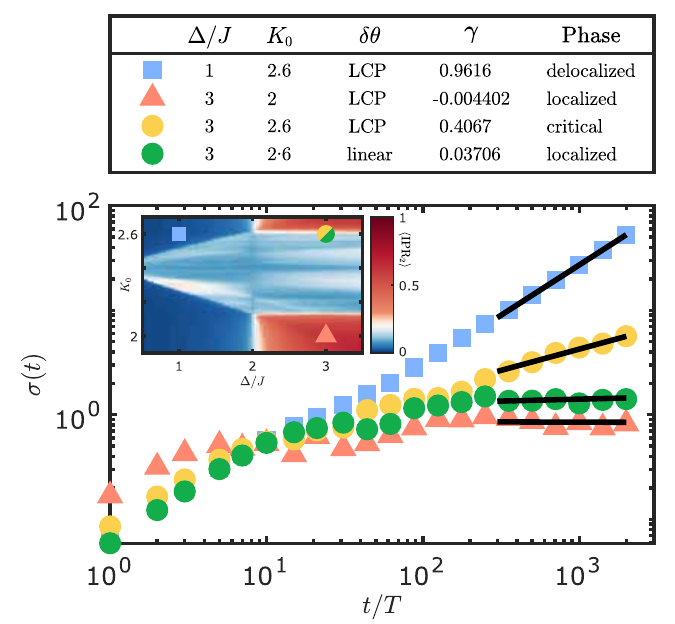}
    \caption{\ybk{Time evolution of the rms width (Eq. \ref{eq:widthExponent}) at different locations on the phase diagram in the space of drive polarization. The inset shows the phase diagram under circularly polarized modulation with markers at the locations where the numerical simulations are conducted. The top panel shows the parameters, polarization, the expansion exponent $\gamma$ and the underlying phases. The bottom panel shows the time evolution of the width $\sigma(t)$ from which the exponents $\gamma$ are obtained by fitting, shown by the black lines.}
    The system is driven for $2000$ cycles at $\hbar \omega = 15J$ with $\varphi_0 = K_0$ and data in each case are averaged over 10 randomly selected values of $\kappa$. The initial state is a single-site excitation $\psi(t) = \delta_{j,0}$ and the Hamiltonian (\ref{DAAH1}) is numerically integrated with the second order split-operator method with a time step of $dt \approx 1.26\times 10^{-5} \hbar/J$.}
    \label{fig:exponents}    
\end{figure}

Experimentally, a shaken bichromatic optical lattice can be implemented either by moving mirrors \cite{zenesini_coherent_2009_DL_Coherent, Rajagopal_phasonic_PhysRevLett.123.223201}, or by relative frequency modulation of counterpropagating lattice beams using acousto-optical modulators \cite{lignier_dynamical_2007_DLExp, shimasaki2023reversible}. We focus on the latter approach and provide a straightforward recipe for achieving circularly polarized modulation. 

Circularly polarized modulation can be achieved most easily by shaking the primary lattice while keeping the secondary lattice static ($x_S (t) = 0$) in the lab frame by introducing a frequency difference $f_P (t) = f_{0,P}\sin(\omega t)$ between the two counterpropagating lattice beams that form the primary lattice. In the co-moving frame, the dipolar and phasonic modulations are then given by \cite{lignier_dynamical_2007_DLExp}
\begin{align*}
    K_0(t) &= \frac{M \pi^2}{\hbar k_P^2} f_{0,P} \sin\left(\omega t\right), \\
    \varphi(t) &= 2\pi\beta \frac{f_{0,P}}{\omega} \sin\left(\omega t + \frac{\pi}{2} \right).
\end{align*}
The phase difference between the two modulations is $\delta \theta = \pi/2$, and so the modulation is elliptically polarized. The condition of circular polarization is $K_0 = K/\hbar \omega = \varphi_0$, that is, 
\begin{equation}
    \frac{M \pi^2 }{\hbar k_P^2} f_{0,P} = 2\pi\beta \frac{f_{0,P}}{\omega} \rightarrow \hbar \omega_\mathrm{RCP} = \frac{4\beta}{\pi} E_\mathrm{r,P}.
\end{equation}
For $\beta = \beta_\mathrm{GR}$, $4\beta_\mathrm{GR}/\pi \approx 0.787$. Thus, when the driving frequency matches the recoil energy with the above relation, the modulation becomes RCP, and the modulation frequency $\omega_\mathrm{RCP}$ is within the first band gap of the optical lattice. In fact, this shaking protocol has already been demonstrated experimentally in recent work \cite{shimasaki2023reversible}, although that work did not focus on critical or topological dynamics. 

Lastly, we comment on experimental signatures of a critical phase. Possibilities include, for example, the imbalance \cite{Luschen_SPME_PhysRevLett.120.160404} or the dynamic exponent of the root-mean-squared (rms) width $\sigma(t)$ of the trapped gas \cite{Ketzmerick_spreading_PhysRevLett.79.1959, shimasaki2022anomalous}. Considering the latter in more detail, the rms width is defined as 
\begin{equation}
    \sigma (t) = \sqrt{\sum_j (j-j_0)^2 \vert \psi_j (t) \vert^2} \sim t^\gamma, \label{eq:widthExponent}
\end{equation}
where $j_0:=\sum_j j \vert \psi_j (t) \vert^2$ is the center of the wave function. The long-time evolution of the rms width, governed by the exponent $\gamma$, can distinguish system's localization properties. For a localized (delocalized) phase, $\gamma = 0 \, (1)$. If the underlying spectrum and the associated eigenstates are multifractal, the dynamic exponent $\gamma$ is in between $0$ and $1$ \cite{Ketzmerick_spreading_PhysRevLett.79.1959}. Thus, we can distinguish these phases by monitoring the expansion of the wave function and extracting the dynamic exponent. 

We theoretically demonstrate the potential use of this experimental signature by computing the time evolution of the wavefunction at exemplary points in the phase diagram, using the Hamiltonian in the length gauge (Eq.~\ref{DAAH1}). The results in Fig. \ref{fig:exponents} show the expected ballistic expansion in the delocalized phase ($\gamma \approx 1$), the suppression of transport at long times in the localized phase ($\gamma \approx 0$), and the critical expansion with $\gamma \approx 0.4067$ when  circularly polarized modulation is applied. In contrast, when the modulation is linearly polarized, the same values of $\Delta$ and $K_0$ no longer lead to critical dynamics but rather to localization; this provides a particularly clean demonstration of control of the state of matter using drive polarization.  

\section{Summary and Outlook} \label{sec:conclusion}
In this work, we introduced combined dipolar and phasonic modulations to the AAH model, and outlined a mapping to a 2D integer quantum Hall system illuminated by light of variable polarization. The combined modulations offer a remarkable degree of control over localization. Phenomena that can thereby be physically realized include tessellated phase diagrams with interlaced localized and delocalized phases, effective time-reversal dynamics, Floquet-engineered multifractality without fine-tuning, and (as described in appendix D) topological edge states.  

These results pave the way for future work, in part because the simplicity of our model lowers the barrier to experimental realization. Here we outline some future opportunities. 
 
A straightforward extension is to use low-dimensional quasiperiodic systems to study the dynamics of corresponding strongly-driven integer quantum Hall systems in higher dimensions \cite{Kolovsky_driven_Harper_PhysRevB.86.054306, Kolovsky_cyclotron_bloch_I_PhysRevE.83.041123, Kolovsky_cyclotron_bloch_II_PhysRevE.86.041146, Kraus_2D4DQH_PhysRevLett.111.226401, Lohse2018_2D4DQH_Exp}. For example, one can generalize our results to study 4D driven quantum Hall systems with 2D doubly-driven bichromatic lattices \cite{Kraus_2D4DQH_PhysRevLett.111.226401, Lohse2018_2D4DQH_Exp}. Extensions to a driven Hofstadter model with a non-abelian gauge field are also possible by considering spinful fermions in the DDAAH model \cite{guan_nonabelian_PhysRevA.108.033305}. Our results may have implications for driven strained Moir\'e superlattices in a magnetic field, where the anisotropic conductivity can be switched by irradiation (Fig.~\ref{fig:cascadeLocTransition} and \cite{Barelli_PhysRevLett.83.5082, Paul_directional_PhysRevLett.132.246402}). \ybk{While we focus on off-resonant driving in this work, resonant driving could allow us to study quantum geometry of a higher-dimensional system using its lower-dimensional counterpart. Possible targets of study include circular dichroism in one dimension \cite{Tran_Dauphin_Grushin_Zoller_Goldman_2017} and quantum metrics \cite{Ozawa_QG_PhysRevB.97.201117}. }  
Finally, the inclusion of interparticle interactions poses a challenging and fundamental question: can one expect a Floquet many-body critical phase? In some static systems, critical phases have been shown numerically to retain their multifractal characteristics in the presence of interactions \cite{Liu_LSP_10.21468/SciPostPhys.12.1.027, Wang_MBC_PhysRevLett.126.080602}, thus appearing as an additional possibility alongside thermalization and many-body localization. Probing the stability of the Floquet critical phase in the presence of interactions is thus a natural target for future investigation \cite{Lazarides_MEMBL_Heating_PhysRevLett.115.030402, Anisimovas_interaction_PhysRevB.91.245135}. 

\begin{acknowledgments}
We acknowledge helpful discussions with Toshihiko Shimasaki, Anna Dardia, Peter Dotti, Chenhao Jin, and Nisarga Paul. We further acknowledge Toshihiko Shimasaki, Anna Dardia, Peter Dotti, and  Jeremy Tanlimco for critical readings of the manuscript.
Use was made of computational facilities purchased with funds from the National Science Foundation (CNS-1725797) and administered by the Center for Scientific Computing (CSC). The CSC is supported by the California NanoSystems Institute and the Materials Research Science and Engineering Center (MRSEC; NSF DMR 2308708) at UC Santa Barbara.
We acknowledge research  support from the Air Force Office of Scientific Research (FA9550-20-1-0240), the NSF QLCI program (OMA-2016245), and the UC Santa Barbara NSF Quantum Foundry funded via the Q-AMASE-i program under Grant DMR1906325. 
The development of the bichromatic optical lattice techniques which are the basis for this work was supported by the U.S. Department of Energy, Office of Science, National Quantum Information Science Research Centers, Quantum Science Center.

\end{acknowledgments}

\appendix

\section{Floquet analysis} \label{apd:FlqAnalysis}
By the Floquet theorem, for a time-periodic Hamiltonian $\hat H(t) = \hat H(t+T)$ there exists a complete and orthornormal basis whose elements $\vert \psi_n (t)\rangle$ are referred to as the Floquet eigenstates which time-evolve as $\vert \psi_n (t+T)\rangle = e^{-i \varepsilon_n T/\hbar} \vert \psi_n (t)\rangle$ where $\varepsilon_n$ is the quasienergy. In analogy to the case of Bloch's theorem, the Floquet eigenstates can be decomposed into a ``plane wave" factor $e^{-i\varepsilon_n t/\hbar}$ and a time-periodic Floquet function $\vert \Psi_n (t)\rangle$: 
\[
\vert \psi_n (t)\rangle = e^{-i\varepsilon_n t/\hbar} \vert \Psi_n (t)\rangle , \quad  \vert \Psi_n (t+T)\rangle = \vert \Psi_n (t)\rangle. 
\]
Notably, the Floquet function can then be represented by a discrete Fourier series as $\vert \Psi_n (t)\rangle = \sum_{m=-\infty}^\infty \vert \phi_{n,m} \rangle e^{im\omega t}$ where $m$ labels the Fourier harmonics. Employing the Fourier decomposition of the time-dependent Hamiltonian  $\hat H(t) = \sum_{m=-\infty}^\infty \hat H_m e^{im\omega t}$, we can derive the following equation from the Schr\"odinger equation: 
\[
\left(\varepsilon_n + m\hbar \omega \right) \vert \phi_{n,m} \rangle = \sum_{m'}\hat H_{m-m'} \vert \phi_{n,m'} \rangle.
\]
This algebraic equation can be written in the following block matrix form, which we refer to as the \textit{Floquet block Hamiltonian} in the main text:  
\begin{widetext}
    \begin{equation}
        \mathcal{H}\varphi_n = \varepsilon_n \varphi_n, \quad \mathcal{H} = \begin{pmatrix}
        \ddots & \hat H_{-1} & \hat H_{-2} & \hat H_{-3} \\
        \hat H_{1} & \hat H_0-m\hbar \omega & \hat H_{-1} & \hat H_{-2} \\
        \hat H_{2} & \hat H_{1} & \hat H_0 - (m+1)\hbar \omega & \hat H_{-1} \\
        \hat H_3 & \hat H_2 & \hat H_1 & \ddots
        \end{pmatrix}, \quad \varphi_n = \begin{pmatrix}
        \vdots \\
        \vert \phi_{n,m} \rangle \\
        \vert \phi_{n,m+1} \rangle \\
        \vdots
        \end{pmatrix} \label{eq:flq_blockH}
    \end{equation}
\end{widetext}
The quasienergy spectrum and properties of Floquet eigenstates such as $\mathrm{IPR}_2^{(i)}$ can thus be extracted by numerical diagonalization with some finite Fourier index $m_\mathrm{cutoff}$ and reconstructed by an inverse discrete Fourier transform. 

In our model, via the Jacobi-Anger expansion, 
\[
e^{i x \sin \theta} = \sum_{m = -\infty}^\infty \mathcal{J}_m (x) e^{im\theta},
\]
the Fourier coefficients $\hat H_m$ of the DDAAH Hamiltonian (\ref{DAAH2}) are 
\begin{align*}
    \hat H_m = \sum_{j=1}^{L-1} &-J e^{im \theta_x} \mathcal{J}_m (K_0)\left((-1)^m\hat c_{j+1}^\dagger \hat c_j + \hat c_{j}^\dagger \hat c_{j+1}\right) \\
    &+ \sum_{j=1}^{L} \Delta e^{im \theta_y} \mathcal{J}_m (\varphi_0)i^m \cos \left(2\pi\beta j +\kappa - \frac{m\pi}{2} \right)\hat n_j.
\end{align*}

\section{High-frequency expansion of modulations with a generic polarization} \label{apd:HFE_generic}

We have reported $\hat H^{(0)}$ and $\hat H^{(1)}$ in the special case of circularly polarized modulation ($K_0 = \varphi_0, \, \delta\theta = \pm\pi/2$). For an arbitrary polarization, the term $\hat H^{(1)}$ can be evaluated as
\begin{widetext}
    \begin{multline*}
        \hat H^{(1)} = \frac{4J\Delta}{\hbar \omega} \sin(\pi\beta) \sum_{m=1}^\infty \frac{i^{m+1}}{m}\sin\left(m(\theta_y - \theta_x)\right)\mathcal{J}_m(K_0)\mathcal{J}_m (\varphi_0) \\
        \times \sum_{j=1}^{L-1}\sin\left(2\pi\beta j + \kappa + \pi\beta -\frac{m\pi}{2}\right) \left( -\hat c_{j+1}^\dagger \hat c_j +(-1)^m \hat c_{j}^\dagger \hat c_{j+1} \right).
    \end{multline*}
\end{widetext}
The interference term $\sin\left(m(\theta_y - \theta_x)\right)$ provides a selection rule of virtual $m$-photon absorption and emission processes. We can see that $\hat H^{(1)}$ vanishes in the case of linearly polarized modulation $\theta_y = \theta_x$, and that only terms with odd $m$ are non-vanishing when $\delta \theta = \pm \pi/2$ due to the interference.  

We have also computed $\hat H^{(2)}$ analytically and numerically \cite{Goldman_PhysRevX.4.031027}, although the results are too lengthy to be fully displayed here. The overall results are quasiperiodically modulated nearest-neighbor tunneling in 1D and additional quasiperiodic potential terms. 

\section{Comparison of multifractal analyses between the Floquet block Hamiltonian and truncated effective Hamiltonian}\label{apd:D2Flq}
\begin{figure}[htb!]
    \centering
    \includegraphics[width = \linewidth]{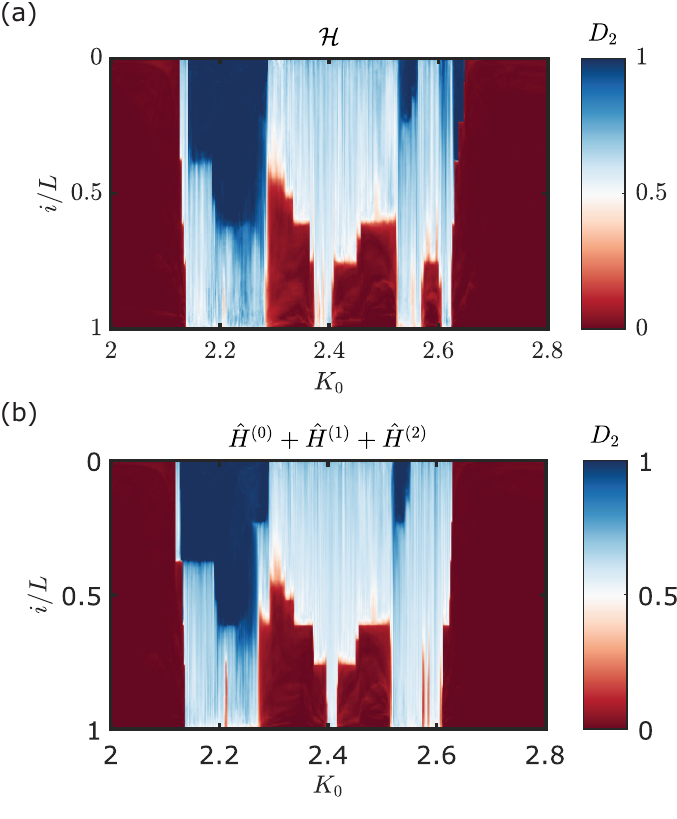}
    \caption{Comparison of calculations of fractal dimensions $D_2$ based on the (a) the Floquet block Hamiltonian (Eq.\ref{eq:flq_blockH}), reproduced from Fig.~\ref{fig:MF_D2}(b) for convenience, and \textbf{(b)} a truncated high-frequency expansion. The data in \textbf{(a)} are reproduced from Fig.\ref{fig:MF_D2}(b) for easier comparison. We recall that the system size is varied from $300$ to $3300$ with $m_\mathrm{cutoff} = 2$ in \textbf{(a)}. In contrast, for the result in \textbf{(b)}, the system size is varied from $300$ to $6000$ with $m_\mathrm{cutoff} = 10$. Both calculations use $\kappa = 0$.}
    \label{fig:D2Floquet}
\end{figure}

In Section \ref{sec:criticalPhase}, we presented the multifractal analysis using the Floquet block Hamiltonian (Eq. \ref{eq:flq_blockH} and Fig. ~\ref{fig:MF_D2}). In principle, this approach works in any frequency regime without the need to evaluate and truncate the high-frequency expansion (Eq. \ref{eq:HFE}). 

However, a realistic numerical study cannot include all the infinitely-many Fourier harmonics, and so there is a cutoff labeled by $m_\mathrm{cutoff}$ in the Fourier components. Thus, this approach may potentially miss the contributions from harmonics higher than $m_\mathrm{cutoff}$ from the Fourier expansion. Nonetheless, their contributions are still relatively small due to two main reasons: the decay of Bessel functions $\mathcal{J}_m(z)$ as a function of $m$ (where $z = K_0$ or $\varphi_0$) at a fixed amplitude $K_0$ or $\varphi_0$, and  Wannier-Stark localization in the Fourier space which describes the wavefunction localized in the Fourier space labeled by $m$, because the diagonal terms in $\mathcal{H}$ proportional to $\omega$ are analogous to a linear potential in the Fourier lattice indexed by $m$ \cite{Lindner_WannierFourierLoc_PhysRevX.7.011018}. 

An associated practical challenge is the computational cost associated with the system-size scaling. In the numerical diagonalization, the number of Floquet-Brillouin zones is $(2m_\mathrm{cutoff}+1)$, and the block matrix has a dimension of $(2m_\mathrm{cutoff}+1)L$ where $L$ is the system size. Consequently, the multifractal analysis with the Floquet block Hamiltonian quickly becomes numerically expensive as the system size $L$ is increased, even for a relatively small $m_\mathrm{cutoff}$. 

Another approach to multifractal analysis is to use the effective Hamiltonian from a truncated high-frequency expansion. Unlike the approach with Floquet block Hamiltonian, this truncation is applied to the expansion index $n$ up to $n_\mathrm{cutoff}$: $\hat H_\mathrm{eff} =\sum_{n=0}^{n_\mathrm{cutoff}} \hat H^{(n)}$. We are thus ignoring contributions that decay as $1/(\hbar\omega)^n$ with $n > n_\mathrm{cutoff}$. Independently, only finitely many Fourier components (up to $m_\mathrm{cutoff}$) can be included. The advantage of this approach is that the dimension of the effective Hamiltonian is the same as that of the original system, $(2m_\mathrm{cutoff}+1)$ times smaller than that of the Floquet block Hamiltonian. Thus, the computational cost of scaling the effective Hamiltonian is much less than that of scaling the Floquet block Hamiltonian, regardless of the $m_\mathrm{cutoff}$ used. Therefore, for the numerical multifractal analysis, the truncated high-frequency expansion is helpful in reaching a larger system size and including more Fourier harmonics, but ignores contributions with strength no larger than $1/(\hbar\omega)^{n_\mathrm{cutoff}}$. These two approaches thus complement each other. 

Here we compare and the contrast these two approaches by numerical calculation of the fractal dimension in the same parameter range. For the truncated high-frequency expansion we take $n_\mathrm{cutoff} = 2$.
The result is shown in Fig. \ref{fig:D2Floquet}. Values of $D_2$ are still non-trivial in a wide range of parameters and for a non-vanishing fraction of eigenstates, and we observe only minor differences between the results of these two calculations. This further supports the existence of the critical phase in our model. 

%% move topological stuff to appendix
\section{From driven Harper-Hofstadter to the generalized Su-Schrieffer-Heeger model} \label{sec:piHHtoSSH}

Although this work primarily focuses on localization properties, drive-induced topological phenomena are also of interest \cite{Oka_Flq_rev_doi:10.1146/annurev-conmatphys-031218-013423, Mei_top_shakenBO_PhysRevA.90.063638}. In this section, we will focus on specific examples where $\beta$ is rational, and the 1D system ceases to be aperiodic and is topologically trivial when static. We will show how combined dipolar and phasonic modulation creates topologically non-trivial edge modes, generalizing previous results such as \cite{Mei_top_shakenBO_PhysRevA.90.063638}, and note an interesting connection between the driven Harper-Hofstadter model and two prototypical models for 1D topological phenomena: the generalized Rice-Mele (RM) and Su-Schrieffer-Heeger (SSH) models \cite{Guo_SSHgeneral_PhysRevB.91.041402}. 

It is known that the 1D RM model and SSH models can be projected from the 2D Harper-Hofstadter model with NNN hoppings \cite{Guo_SSHgeneral_PhysRevB.91.041402, Martinez_SSH3_PhysRevA.99.013833, Ganeshan_majorana_PhysRevLett.110.180403}. Consider the case where the 2D Harper-Hofstadter model with NNN hoppings has a magnetic flux $\beta = 1/q$ (where $q$ is a positive integer) through one plaquette. The projected 1D model is a superlattice with $q$ sites per unit cell. In the projected space, both the on-site potential and the nearest-neighbor tunneling are sinusoidally modulated with the same spatial period $q$. The projected model then describes a RM model with $q$ sites per unit cell. When the on-site potential vanishes, this becomes the SSH model with $q$ sites per unit cell \cite{Guo_SSHgeneral_PhysRevB.91.041402}. 

As previously described in Section \ref{sec:criticalPhase}, these NNN hoppings can be induced in the ordinary 2D Harper-Hofstadter model by elliptically polarized irradiation. It immediately follows that an effective generalized RM model can be obtained by projection of an elliptically-driven 2D Harper-Hofstadter model. Now we recall that the strength of the on-site potential in the 1D space corresponds to the tunneling strength along the extra dimension in the 2D space (Fig.\ref{fig:schematic}). Thus, if the external driving can make the tunneling along the extra dimension effectively vanish in the 2D space, then the projected 1D model has no on-site potential: it becomes the SSH model with $q$ sites per unit cell. A vanishing tunneling strength along the extra dimension corresponds precisely to dynamic localization in 2D, or phasonically-induced destruction of localization in 1D, when $\varphi_0$ is one of the zeros of $\mathcal{J}_0(\varphi_0)$. 

\begin{figure}[thb!]
    \centering
    \includegraphics[width = \linewidth]{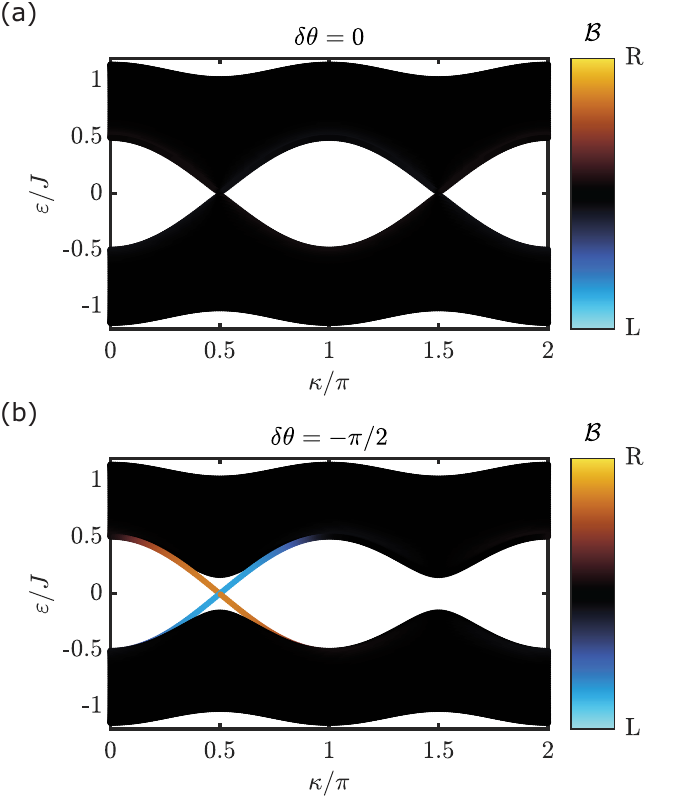}
    \caption{Comparison of the quasienergy spectra of a dimerized lattice $(\beta = 1/2)$ under \textbf{(a)} linearly polarized modulation and \textbf{(b)} circularly polarized modulation that leads to a Floquet Rice-Mele model. \textbf{(a)} Band-touching (Dirac points) can be seen at the phases $\kappa = \pi/2$ and $\kappa = 3\pi/2$. \textbf{(b)} When the modulation is circular, gaps open at the Dirac points and edge states form. The spectra are obtained by diagonalizing the Floquet block Hamiltonian $\mathcal{H}$ with $\lambda = 1$, $K_0 = \varphi_0 = 1.5$ and $\hbar \omega = 15J$. }
    \label{fig:RM}    
\end{figure}

\begin{figure}[thb!]
    \centering
    \includegraphics[width = \linewidth]{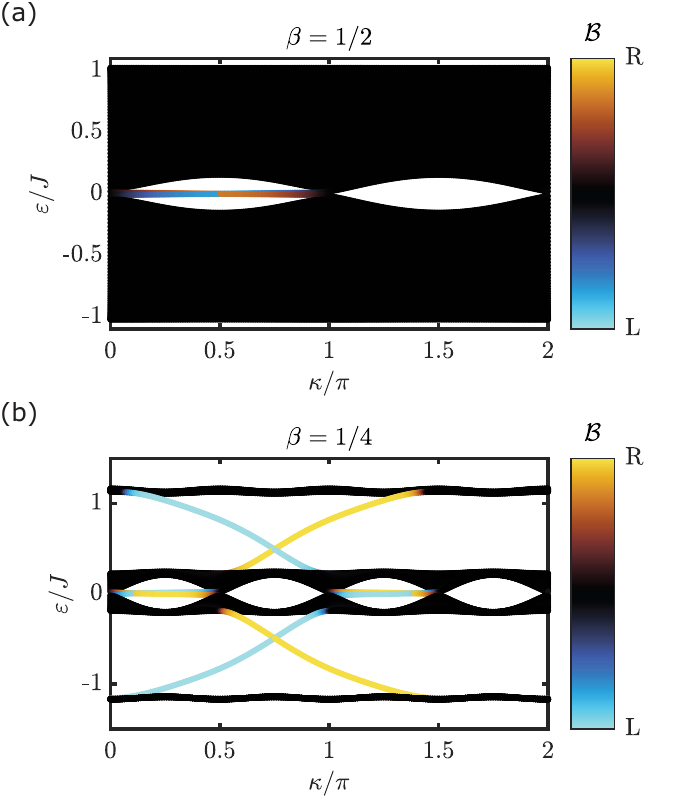}
    \caption{Edge modes of the commensurate DDAAH model under elliptically polarized modulation ($\delta\theta=-\pi/2$) driven at $\varphi_0 \approx 2.4048$, the first zero of $\mathcal{J}_0(\varphi_0)$. \textbf{(a)} Quasienergy spectrum for $\beta = 1/2$ and $\lambda = 1$. Two nearly degenerate edge modes between $0<\kappa < \pi$ are visible around $\varepsilon/J = 0$. \textbf{(b)} Quasienergy spectrum for $\beta = 1/4$ at $\lambda \approx 6.1047$ (so $|J_1/\mathcal{J}_0(K_0)| = 1$). A pair of nearly degenerate edge modes are visible between $0<\kappa < \pi/2$ and $\pi<\kappa<3\pi/2$ inside the gaps around $\varepsilon/J = 0$, and two other quantum-Hall type edge modes also appear. Both calculations are done at $\hbar \omega = 15J$, $L = 100$ and $K_0 = 1.5$ using the Floquet block Hamiltonian (Eq. \ref{eq:flq_blockH}) with open boundary conditions.}
    \label{fig:edgeModeSSH}    
\end{figure}

As an example, we begin with $\beta = 1/2$ in the static AAH model. This static dimerized lattice with spatially homogeneous $J$ is topologically trivial. It is projected from the $\pi-$flux Harper-Hofstadter model which is also topologically trivial due to time-reversal symmetry. Varying $\kappa$, two Dirac points, associated with the $\pi$-flux model, appear at $\kappa = \pi/2$ and $\kappa = 3\pi/2$, as shown in Fig. \ref{fig:RM}(a). When an elliptically polarized modulation ($\delta \theta = \pm \pi/2$) is applied, however, the effective Hamiltonian $\hat H_\mathrm{eff} = \hat H^{(0)}+\hat H^{(1)}$ becomes
\begin{align}
    \hat H_\mathrm{eff} = &-\sum_{j=1}^{L-1}\left(J_\mathrm{eff}(K_0) + (-1)^j J_1(K_0, \varphi_0) \cos \left(\kappa \right) \right)\hat c_{j+1}^\dagger \hat c_j \nonumber \\ 
    & + \mathrm{h.c.} + \Delta_\mathrm{eff}(\varphi_0) \sum_{j=1}^{L}\sin\left(\kappa\right) (-1)^j \hat n_j. 
\end{align}
This is a Floquet RM model \cite{Shen_2012} that exhibits topological pumping as we scan $\kappa = \Omega t$ with $\Omega \ll \omega$. In full accordance with the static RM model, we observe that gaps open at the Dirac points and a pair of edge states localized at each end of the lattice emerge and traverse the gap as shown in Fig. \ref{fig:RM}. These edge states are numerically identified by the edge-locality marker of the $i^\mathrm{th}$ eigenstate \cite{Tran_EdgeLocMarkerPhysRevB.91.085125}
\begin{equation}
    \mathcal{B} = \sum_{j\in\{\mathrm{L},\mathrm{R} \}}|\psi_j^{(i)}|^2 \label{eq:edgeLocMarker}
\end{equation}
where L and R represent sites at left and the right edge, respectively. 

This observation is analogous to the case of irradiated graphene, where gaps open at the Dirac points due to drive-induced NNN hoppings \cite{Kitagawa_PhysRevB.84.235108}. Unlike the case of eigenstate multifractality, here the higher-order terms in the high-frequency expansion do not destroy edge states as they are topologically robust. We note that the same model at $\beta = 1/2$ has been studied before, and identified as a Floquet topological insulator \cite{Mei_top_shakenBO_PhysRevA.90.063638}. 

The RM model becomes the celebrated SSH model when the on-site potential vanishes. In our system, this condition corresponds to $\Delta_\mathrm{eff} (\varphi_0) = 0$ due to an appropriately chosen phasonic component of the elliptically polarized modulation. From the 2D perspective, $\Delta_\mathrm{eff} (\varphi_0) = 0$ corresponds to the condition of dynamic localization along the extra dimension. The effective Hamiltonian then becomes
\begin{align}
    \hat H_\mathrm{eff} \approx -\sum_{j=1}^{L-1} &\left(J_\mathrm{eff} + (-1)^j J_1(K_0, \varphi_0)\cos \left(\kappa \right) \right) \hat c_{j+1}^\dagger \hat c_j \nonumber \\
    &+ \mathrm{h.c.}.
\end{align}
This (Floquet) SSH model can be transformed into two decoupled chains of Majorana fermions, and features degenerate zero-modes across the gap when $0<\kappa<\pi$ \cite{Ganeshan_majorana_PhysRevLett.110.180403, Shi_MZMExpAAH_PhysRevLett.131.080401}. Our numerical calculations from the Floquet block Hamiltonian are in agreement with this prediction, as shown in Fig.~\ref{fig:edgeModeSSH}. Higher-order terms in the high frequency expansion couple the two independent Majorana chains, lifting the degeneracy of the zero-modes \cite{Ganeshan_majorana_PhysRevLett.110.180403}, which can be observed in Fig.~\ref{fig:edgeModeSSH}(a). However, their appearance and locations are robust against such perturbations. 

As discussed earlier, the analyses above can be generalized to other rational $\beta=1/q$ where $q$ is a positive integer \cite{Guo_SSHgeneral_PhysRevB.91.041402}. The projected 1D system will then have $q$ lattice sites per unit cell. For example, the case $\beta = 1/4$ was first studied by \cite{Ganeshan_majorana_PhysRevLett.110.180403}. When there is no modulation, the system is again topologically trivial in 1D. Topological edge states show up once we apply elliptical modulations at $\Delta_\mathrm{eff}(\varphi_0) = 0$, so that the effective Hamiltonian becomes the SSH4 model in 1D. Apart from the two pairs of  quantum Hall edge states, the quasienergy spectrum shown in Fig.\ref{fig:edgeModeSSH}(b) again contains pairs of nearly-degenerate edge states as predicted by~\cite{Ganeshan_majorana_PhysRevLett.110.180403}.

\bibliography{refs}

\end{document}